\documentclass[onecolumn,fleqn,prd,aps,showpacs,preprintnumbers,nofootinbib,eqsecnum]{revtex4}
\usepackage{graphics,graphicx}
\usepackage{amsmath}
\usepackage{amsfonts}
\usepackage{mathrsfs}
\usepackage{amsbsy}
\usepackage{calrsfs}
\usepackage{caption}
\DeclareMathAlphabet{\mathcalligra}{T1}{calligra}{m}{n}
\usepackage{cancel}

\usepackage{psfrag}
\usepackage{tabularx}
\usepackage{dcolumn}
\usepackage{color}
\usepackage{fancybox}
\usepackage{ulem}
\usepackage{xcolor}
\xdefinecolor{mylinkcolor}{rgb}{0,0,0.5}
\usepackage[
    bookmarksnumbered, bookmarksopen, bookmarksopenlevel=2,
	breaklinks=true,
	colorlinks=true, filecolor=mylinkcolor, citecolor=mylinkcolor,
	linkcolor=mylinkcolor, urlcolor=mylinkcolor, menucolor=mylinkcolor,
]{hyperref}

\usepackage[T1]{fontenc}
\usepackage{xcolor}
\usepackage{tikz}
\usepackage{sidecap}
\usepackage{fancybox}
\usepackage{dsfont}

\usepackage{amsmath}
\usepackage{amsfonts}
\usepackage{amssymb}
\allowdisplaybreaks

\def\vct#1{{\mathbf{#1}}}

\def\neanl{\nonumber\\}

\newcommand{\HAM}[2]{	{H}_{\rm {#1}}^{\rm {#2}}}

\newcommand{\vSone}{ \vct{S}_1 }
\newcommand{\vStwo}{ \vct{S}_2 }
\newcommand{\cInv}[1]{ c^{-#1}}

\newcommand{\LSOne}{ \scpm{\vAng}{\vSone} }
\newcommand{\LStwo}{ \scpm{\vAng}{\vStwo} }

\newcommand{\JLSGSum}{ \left( J^2-\Ang^2-\SpinTot^2 \right) }
\newcommand{\SpinTot}{ S_G }
\newcommand{\phiSpin}{ \phi_S}

\newcommand{\vnunit}{{\vnxa{12}}}

\newcommand{\vnxa}[1]{{\vct{n}}_{#1}}

\newcommand{\scpm}[2]{(#1 \cdot #2)}

\newcommand{\rel}{r}

\newcommand{\order}[2]{{\cal O}({#1}^{#2})}
\newcommand{\MeAno}{{\cal M}}
\newcommand{\MeMo}{{\cal N}}

\newcommand{\EccAno}{{\cal E}}
\newcommand{\NRG}{|E|}

\newcommand{\ang}{L}
\newcommand{\vAng}{\vct{L}}
\newcommand{\Ang}{L}

\renewcommand{\em  }[1]{{\sl{ #1}}}
\renewcommand{\emph}[1]{{\sl{ #1}}}

\newcommand{\pert}{ {\rm pert} }
\newcommand{\Hint}{H^*}
\newcommand{\Hpert}{H^\pert}

\begin{document}
\allowdisplaybreaks

\title{Eccentric Motion of Spinning Compact Binaries}
\author{Manuel Tessmer}
\email{M.Tessmer@uni-jena.de}
\affiliation{Institut f\"ur Angewandte Physik,
	Friedrich--Schiller--Universit\"at,
	Albert-Einstein-Stra{\ss}e 15,
    07745 Jena, Germany, EU}
\author{Gerhard Sch\"afer}
\email{Gerhard.Schaefer@uni-jena.de}
\affiliation{Theoretisch--Physikalisches Institut,
	Friedrich--Schiller--Universit\"at,
	Max--Wien--Platz 1, 07743 Jena, Germany, EU}
\date{\today}

\begin{abstract}
The equations of motion for spinning compact binaries on eccentric orbits
are treated perturbatively in powers of a fractional mass-difference ordering 
parameter. The solution is valid through first order in the mass-difference
parameter. A canonical point transformation 
removes the leading order terms of the spin-orbit Hamiltonian which induce 
a wiggling precession of the orbital angular momentum around the conserved total angular momentum, 
a precession which disappears in the case of equal masses or one single spin. Action-angle 
variables are applied which make a canonical perturbation theory easily treatable.
\end{abstract}
\maketitle

\setlength\mathindent{1pt}

\tableofcontents

\section{Introduction}

Compact binary star systems are often investigated in general relativity when moving on orbits with zero eccentricity.
This is usually justified by the circularisation effect due to
radiation reaction for \em{isolated} binaries 
\cite{Peters:Mathews:1963, Peters:1964, Gopakumar:Iyer:1997, Arun:Blanchet:Iyer:Qusailah:2008,
Arun:Blanchet:Iyer:Qusailah:2008-1, Arun:Blanchet:Iyer:Sinha:2009}
which becomes rather strong in the late stages of the binary's life.
For this reason, numerical relativity simulations of compact binaries
(which typically model the late inspiral phase) often start from
quasi-circular orbits \cite{Pfeiffer:2012}. Nonetheless, binaries can retain finite eccentricity through
various mechanisms involving either additional bodies or gaseous
environments \cite{OLeary:Kocsis:Loeb:2009}. Eccentric binaries typically lead to enhanced and
more complex gravitational wave (GW) emission compared to the
quasi-circular case \cite{Yunes:Arun:Berti:Will:2009, Gold:Brugmann:2013} which leads to two consequences: (i)
Eccentric binaries can be detected out to larger distances (up to two
orders of magnitude in detection volume for adLIGO
\cite{Yunes:Arun:Berti:Will:2009})
than quasi-circular binaries (everything else being equal), which affects
their (poorly known) event rates \cite{OLeary:Kocsis:Loeb:2009}. (ii) Parameter estimation for
GW detectors typically adopt quasi-circular templates which can
severely limit the ability to detect GWs and to recover source
parameters \cite{Brown:Zimmerman:2010, Yunes:Arun:Berti:Will:2009}.

In order to also model the GW signals from \em{eccentric} binaries
reliably, it is essential to include higher order general relativistic effects that
are very well described within the post-Newtonian framework,
also see the discussion in Ref. \cite{Nielsen:2012}.
The reason for it is that spin precession and periastron advance
will generate modifications (e.g. side-bands) in the GW Fourier
domain. If they are not included, the correlation of real detector 
data with incomplete and therefore non-optimal GW templates leads to
computed system parameters that are displaced with respect to "real" ones,
although the signal may be covered more or less effectively for special
configurations. 

The point-mass contributions to the post-Newtonian Hamiltonians in the Arnowitt-Deser-Misner
(ADM) gauge have been computed through the fourth post-Newtonian (4PN) order 
(=$\order{ {v/c} }{}^8$, where
$v$ is a typical internal velocity and $c$ the speed of light) in 
Ref.~\cite{Damour:Jaranowski:Schafer:2014} (the 3PN calculation
can be found in an earlier publication \cite{Damour:Jaranowski:Schafer:2001}).
The spin-orbit contributions are derived
in Refs. \cite{Damour:Jaranowski:Schafer:2008:1} and \cite{Damour:Jaranowski:Schafer:2008:3}
through next-to-leading order (=:NLO) and 
in Ref.~\cite{Hartung:Steinhoff:2011:1}
through next-to-next-to-leading order (=:NNLO) for compact binaries. 
NNLO effects in the spin-orbit coupling for an arbitrary number $n$ of
compact spinning objects have been derived in \cite{Hartung:Steinhoff:2010}.

The Hamiltonian prescription leads to a number of evolution equations
for the radial part of the binary motion and for quantities being related
to the orientations of the spins and the orbital plane. The solution to
the linear-in-spin problem for compact binary systems has been discussed
frequently in the recent years, see below.
For example, in
\cite{Tessmer:Steinhoff:Schafer:2013},
the circular-orbit motion has been solved with the help
of a sequence of Lie transformations.
In other publications, see \cite{Keresztes:Mikoczi:Gergely:2005},
a Kepler equation for compact binaries with spin has been given;
in \cite{Majar:Vasuth:2008} the GW forms of eccentric
binaries with spin were worked out -- the equations of motion of the
entering spin orientation angles are given, but not solved there.
In \cite{Blanchet:Buonanno:Faye:2011}, tail-induced spin-orbit
effects in the energy flux and the GW forms have been derived for
circular orbits with arbitrary masses. 

Summarising, in the current article we generalise the recent analytic results
that are known (we omit the included spin and PN orders here), for
\begin{itemize}
\item circular orbits, arbitrary spins and masses 
\cite{Apostolatos:1995, Blanchet:Buonanno:Faye:2006,
Blanchet:Buonanno:Faye:2006:err,
Blanchet:Buonanno:Faye:2006:err:2,
Faye:Blanchet:Buonanno:2006,
Arun:Buonanno:Faye:Ochsner:2009},
\item eccentric orbits, arbitrary spins, but equal masses \cite{Konigsdorffer:Gopakumar:2005},
\item eccentric orbits, one single spin, and arbitrary masses \cite{Konigsdorffer:Gopakumar:2005},
\item eccentric orbits, aligned spins and arbitrary masses \cite{Tessmer:Hartung:Schafer:2010, Tessmer:Hartung:Schafer:2012},
\item circular  orbits, arbitrary spin orientations, allowing slightly unequal masses \cite{Tessmer:2009, Tessmer:Steinhoff:Schafer:2013},
\end{itemize}
and references therein,
to 
{\bf eccentric orbits, arbitrary spins and allowing slightly unequal masses}.
For a first insight, we will include the gravitational leading-order spin-orbit
coupling and the second post-Newtonian (2PN)
point-mass (PM) interaction Hamiltonian for compact binaries
\cite{Damour:Schafer:1985,Schafer:1985}.

Spin(a)-spin(b) and spin-squared couplings also have to be included at some
instant of time in their orbital evolution, especially at the late stage of the
inspiral, but we disregard them in this article for two reasons. The first one
is that they turn out to be small compared to the other terms considered here,
at least at large binary separations where the orbital angular momentum is much
larger than the spin. The second reason is that they cannot be regarded
as a contribution that is growing with the mass difference and, for equal masses, 
a closed-form solution for precession including those terms is not known
until now.

Our tool will be the application of action and angle (AA) variables
(see our subsection \ref{SubSec::AAVarsBasics} and Refs. \cite{Sommerfeld:1951, Alexander:1987}
for their definitions and applications)
and a subsequent generalisation of the Delaunay variables
(see e.g. \cite{Vinti:1998}) which are derived from the AA variables.%
\footnote{Those are related to the recently introduced ``Hill - inspired'' variables 
(see \cite{Hergt:Shah:Schafer:2013} for reasons of this terminology)
for compact binaries with spin.}
The generalisation will be performed in three steps:
(i) taking the expression of the interaction Hamiltonian in terms of the
variables provided in \cite{Tessmer:Steinhoff:Schafer:2013},
(ii) computing the action-angle variables 
from those expressions, and
(iii) eliminating the degeneracy at Newtonian level, leading to the Delaunay-type
variables with spin, using the definition $\vAng=\vct{\rel}\times\vct{p}$.%
\footnote{This canonical definition was not used within the Hill variables, see \cite{Mai:Schneider:Cui:2008}.
For reasons of current research in a slightly different context, also concerning the discussion in a current
article \cite{Gupta:Gopakumar:2013}, we like to give reference to the publication
of Gurfil et al. \cite{Gurfil:Elipe:Tangren:Efroimsky:2007}, dealing with
a distinction of the usage of $\vct{L}_{N} := \vct{\rel} \times \vct{\dot{\rel}}$
which is not a canonical quantity.
Note that the used variables do \em{not} diverge for small azimuthal
angles $\Theta$ as stated by those authors -- in fact, they degenerate
in the exact $\Theta = 0$ case.}
In this context we like to mention a perturbative treatment of star resonances 
in Newtonian orbits by \cite{Alexander:1987}, where also action-angle variables 
came to application to characterise the zero-order problem where no oscillations 
take place.

At this point, we like to state why we prefer to work with canonical variables.
Their advantage is to make canonical perturbation theory
feasible. Although it is not a necessity to tackle perturbation problems in this way
{\cite{Cary:Littlejohn:1983}}, 
it makes the calculation more practical
because standard Poisson brackets remain valid to obtain the EOM (equations of motion)
after any transformation.

The article is organised as follows.
In Section \ref{Sec::Interactions} we provide
the Hamiltonian interaction terms.
In Section \ref{Sec::SolutionEqualMasses} we discuss
the known solution to the problem of unperturbed Hamiltonian
equations of motion for eccentric binaries with spin-orbit interaction.
In Section \ref{Sec::HJ-Theory} we summarise
the main aspects of the Hamilton-Jacobi (HJ) theory
which is used to solve the perturbed equations of motion
in a specific manner. We summarise the definition
of action-angle variables for librational
motion and we present some techniques for the radial
part of the generating function and also those
for the elimination of degeneracy conditions
in the resulting equations of motion.
In a subsequent Section
\ref{Sec::HJ_Perturbation_Theory} we present our main result: 
the application of the HJ perturbation theory to the
case of eccentric binaries with spin, where we
expand the solution to the first order of the
mass difference function $\epsilon$.

\section{Included Interaction Terms}
\label{Sec::Interactions}

The point-mass Hamiltonian (subscript ``PM'') to second post-Newtonian accuracy
will be given below. The symbols are explained in Table~\ref{Tab::Shorthands}.
We work in dimension-less units as given in Eqs.~(6)-(9) of \cite{Tessmer:Hartung:Schafer:2010}
to obtain Eqs. (\ref{Eq::Ham_PM_Newt}--\ref{Eq::Ham_LO_SO}) below,
with the only exception of additionally imposing fast-spinning components, for convenience of 
the reader\footnote{The LO SO interaction is formally of 1PN order.
Imposing fast-spinning components, it is shifted to 1.5PN order, slow rotation shifts it further to 2PN order.}.
We set c=1, but retain c as a power-counting parameter in order
to easily plug in numbers for explicit examples.
\begin{eqnarray}
\label{Eq::Ham_PM_Newt}
 \HAM{PM}{N}
&=&
\frac{p_r^2}{2}+
\frac{\ang^2}{2 r^2}-\frac{1}{r}
\,, \\
 \HAM{PM}{1PN}
&=&
\cInv{2} \left\{
   \frac{{\ang}^4 (3 \eta -1)}{8 r^4}-\frac{{\ang}^2
   (\eta +3)}{2 r^3}+\frac{1}{2 r^2}
+
p_r^2 \left(\frac{(3 \eta -1) L^2}{4 r^2}-\frac{2 \eta +3}{2 r}\right)+\frac{1}{8} (3 \eta
   -1) p_r^4
\right\}
\,, \\
 \HAM{PM}{2PN}
&=&
\cInv{4} \biggl\{
   \frac{\ang^6 (5 (\eta -1) \eta +1)}{16
   r^6}-\frac{\ang^4 \left(3 \eta ^2+20 \eta -5\right)}{8
   r^5}+\frac{\ang^2 (8 \eta +5)}{2 r^4}+\frac{-3 \eta -1}{4 r^3}
\neanl &+& 
 p_r^6\frac{1}{16} (5 (\eta -1) \eta +1)
+p_r^4 \left(\frac{-8 \eta ^2-20 \eta +5}{8 r}+\frac{3 (5 (\eta -1) \eta +1) L^2}{16
   r^2}\right)
\neanl &+& 
 p_r^2 \left(\frac{\left(-4 \eta ^2-20 \eta +5\right) L^2}{4 r^3}+\frac{3 (5
   (\eta -1) \eta +1) L^4}{16 r^4}+\frac{11 \eta +5}{2 r^2}\right)
\biggr\}
\,.
\end{eqnarray}
The terms linear in spin through leading order \cite{Damour:Jaranowski:Schafer:2008:1}
read, ``SO'' denoting spin-orbit coupling and ``LO'' leading order, 
\begin{eqnarray}
\label{Eq::Ham_LO_SO}
\HAM{LO}{SO} &=&  \frac{\cInv{3}}{4 r^3}
 \left\{ 
 \left( 2 \eta +3 \sqrt{1-4 \eta }+3 \right) {\LSOne}
+\left( 2 \eta -3 \sqrt{1-4 \eta }+3 \right) {\LStwo}
 \right\} \,.
\end{eqnarray}
Those Hamiltonians generate equations of motion that, currently,
can be solved only in a perturbative manner. 
One can construct a more practical set of spin variables that
distinguish ``constant'' from ``oscillatory'' (the term ``constant'' is equal to
{``integral of motion''} and ``oscillatory'' is equal to ``give zero time average'';
both are meant in a context that we will
explain later on in Subsec. \ref{SubSec::SeparatingAAVars}) contributions.
We will give the Hamiltonian in these new coordinates
in Section~\ref{Sec::HJ-Theory}.
Let us first turn to the known solution for binaries of
equal masses (also including the single-spin case) 
-- which will serve as a basis for our 
calculation.

\section{Solution to the eccentric spin-orbit problem at leading order with equal masses}
\label{Sec::SolutionEqualMasses}

We define the orbital plane to be that plane which is moving perpendicularly
to the canonical orbital angular momentum $\vAng$.
The motion of compact binaries in the orbital plane can be prescribed by
the following system of equations, which uses definitions of several
orbital elements to be found in Table~\ref{Tab::Shorthands}:
\begin{subequations}
\label{Eq::QKP_EqualMass}
\begin{align}
 r	&= a_r \left(1-e_r \cos \EccAno \right) \,,\\
 \phi	&= 2 \arctan \left\{ \sqrt{\frac{1+e_\phi}{1-e_\phi}}
	      \tan{\frac{\EccAno}{2}}\right\} + \order{c}{-4} \,,\\
 \MeAno &= \EccAno - e_t \sin \EccAno + \order{c}{-4} \,.
\end{align} 
\end{subequations}
\\
The geometrical meaning of the above relations (at Newtonian level)
may be found, for example in Colwell's book~\cite{Colwell:1993}. 
Their derivation is given in, e.\,g., \cite{Memmesheimer:Gopakumar:Schafer:2004} 
and for the aligned-spin case in \cite{Tessmer:Hartung:Schafer:2012}, including
the energy and angular momentum decay due to radiation reaction.
For the leading-order spin-orbit case with (i), single spin or (ii), equal
masses, the above terms get spin-orientation corrections, see
\cite{Konigsdorffer:Gopakumar:2005}, and the following orientation equation
\begin{equation}
\label{Eq::Upsilon_KG04}
 \Upsilon - \Upsilon_0 = \frac{\chi_{so} J}{c^3\Ang^3} (\phi + e \sin \phi) \,,
\end{equation}
has to be added to prescribe the full conservative motion of the system. 
In Eq.~(\ref{Eq::Upsilon_KG04}), $\chi_{so}$ is a term that is either equal to $7/8$ for the equal-mass
case or equal to a function of the masses in the single-spin case 
(see their Eqs.~(2.5a, 2.5b) and (4.36)), the angle $\Upsilon$
is the canonical coordinate conjugate
to the ``momentum'' $J$ and $\phiSpin$ the one associated to $\SpinTot$.
The solution to $\phiSpin$ may also be given, but is irrelevant in that case because it enters nowhere explicitly.
\\ \newline
\begin{center}
\vspace{-0.5cm}
\begin{table}[!h]
\begin{tabular}[c]{ l | r }
\hline
Quantity  & Description \\
\hline \hline
	$c^{-1}$		& \dotfill Power counting for post-Newtonian orders \\
	$n$PN			& \dotfill $n^{\rm th}$ post-Newtonian order,  $\order{c}{-2n}$ \\
	$\eta$			& \dotfill Symmetric mass ratio: $\eta:= m_1 m_2/(m_1+m_2)^2$\\
	$|E|$ 			& \dotfill Absolute value of binding energy 				\\
	$\ang$ 			& \dotfill Angular momentum of orbit, $\ang:=|\vct{\ang}|$		\\
	$\MeAno, \ell_D$	& \dotfill Mean anomaly  						\\
	$\MeMo$			& \dotfill Mean motion or radial angular velocity, respectively		\\
	$\EccAno$		& \dotfill Eccentric anomaly 						\\
	$\vct{p}$		& \dotfill Linear momentum: $p := |\vct{p}|$, $p_\rel:= \scpm{\vnunit}{\vct{p}} $ \\
	$\vct{\rel}$		& \dotfill Radial separation: $\rel := |\vct{\rel}|$, $\vnunit := \vct{\rel}/\rel$ \\
	$\SpinTot$		& \dotfill Total spin: $\SpinTot := |\vSone+\vStwo|$ \\
	$\phi$			& \dotfill Orbital phase, measured from the pericenter			\\
	$\phiSpin$		& \dotfill Spin phase, see Fig.~1 of \cite{Tessmer:2009}		\\
	$\Phi$			& \dotfill Total orbital phase in one radial period				\\
	$\Upsilon$		& \dotfill Rotation angle for $\vAng$ around fixed unit vector $\vct{e}_Z$		\\
	$a_r$			& \dotfill Semimajor axis						\\
	$e_{r}$			& \dotfill Radial eccentricity						\\
	$e_\phi$		& \dotfill Phase eccentricity 						\\
	$e_t$			& \dotfill Time eccentricity 						\\
	$\epsilon$		& \dotfill Mass difference function: $\epsilon^2 := \frac{1}{4}-\eta$	\\
\hline
\end{tabular}
\caption{Shorthands of quantities frequently used in this article.
The Poisson brackets for the coordinates and momenta are to be taken from \cite{Tessmer:Steinhoff:Schafer:2013}.}
\label{Tab::Shorthands}\vspace{-0.5cm}
\end{table}
\end{center}
The quantities $e_r,a_r$ and so on essentially depend on the included
interaction terms. This parameterisation will enter, at its leading order,
the solution to the perturbed motion as basis.
It will be applied to the HJ theory (which is in fact standard), 
summarised in the subsequent section for convenience of the reader.

\section{Hamilton-Jacobi theory}
\label{Sec::HJ-Theory}
The Hamilton-Jacobi theory is often used in celestial mechanics
to transform the considered problem to variables in which the dynamics appear in a much
simpler form compared with the initial one. It is often asked for a
canonical transformation which makes the new momenta
to be constant (maybe constantly equal to zero) and coordinates
which are linear functions of time.
The generating function (let us label it $S$), as it is the case in 
our article, has to be found accordingly in a perturbative manner. It is of
the physical dimension of an action and is, therefore,
simply called the \em{``action''}  in the subsequent lines.

\subsection{Action and angle variables: basics}
\label{SubSec::AAVarsBasics}
In this section we will derive the action-angle variables for the
equal-mass-two-spin system.
Before that, let us state why those kinds of variables are so useful.
Throughout this section, the Einstein summation convention is \em{not employed}.
The starting point is the Hamilton-Jacobi equation
\begin{equation}
 H \left( q, \frac{\partial S}{\partial q} \right) = -\frac{\partial S}{\partial t} \,.
\end{equation}
Let us suppose that the energy, or the value of the Hamiltonian, is conserved.
Then, taking this as input for the HJ equation, the action can be separated as follows:
\begin{equation}
\label{Eq::DefAction}
 S = -E\,t + W(q) \,,
\end{equation}
The function $W(q)$ of spatial coordinates is called the {\em characteristic function}.
There exists a number of conserved quantities $\alpha_1, \alpha_2,...$ -- we may, for
example, define $\alpha_1 = E, \alpha_2=\Ang$... for a system in which the magnitude of the
orbital angular momentum and further momenta are also conserved.
The following quantities, called \em{action variables}, turn out to be interesting
when related to $W$,
\begin{equation}
 J_k := \frac{1}{2\pi}\oint p_k {\rm d} q_k\,,
\end{equation}
where the integral is meant for one complete orbit.
Here, it holds $p_k=p_k(q_i, \alpha_i)$ for Staeckel systems.
The $p_k$ being $q_k$ librational, these $J_k$ do not depend
on $q_k$ any more and thus, one may express the $J_k$ in terms
of the $\alpha_i$ alone,
\begin{equation}
 J_k = J_k \left( \alpha_1, \alpha_2, \dots \right)\,.
\end{equation}
If we turn these relations ``inside-out'', giving $\alpha_i= \alpha_i\left( J_k \right)$,
we obtain for the characteristic function
\begin{equation}
 W = W\left( \boldsymbol{q}, \boldsymbol{J} \right)\,.
\end{equation}
Because the generating function is of that special type and
$S: \left(p, q\right) \rightarrow \left(J,\omega\right)$, where
the $q$ are old coordinates and $J$ are the new momenta,
one computes the coordinate transformation according to
\begin{eqnarray}
 p_k &=& \frac{\partial W}{\partial q_k} \,, \\
 w_k &=& \frac{\partial W}{\partial J_k} \,.
\end{eqnarray}
The Hamiltonian $\HAM{}{}$, as it is conserved and identified with $\alpha_1$, is now a
function of the $J's$ alone,
\begin{equation}
 H = \alpha_1 \left( \boldsymbol{J} \right) \,.
\end{equation}
The main frequencies can be obtained via
\begin{eqnarray}
 \dot{J}_k &=& -\frac{\partial \HAM{}{}}{\partial w_k} = 0 \,, \\
 \dot{w}_k &=&  \frac{\partial \HAM{}{}}{\partial J_k}
	    = \frac{\partial \alpha_1 \left( \boldsymbol{J} \right)}{\partial J_k} =: \nu_k\,.
\end{eqnarray}
In the subsequent lines we will present the calculation
of essential action and angle variables and how to deal
with degenerate systems. In the end, we will perform a
transformation to variables related to the well-known
Delaunay variables.

\subsection{Separating the action: AA-Variables for the integrable system}
\label{SubSec::SeparatingAAVars}
Taking the Hamiltonian in the form of \cite{Tessmer:Steinhoff:Schafer:2013}
and replacing all the momenta (especially: the spins) by derivatives
of the action integral, we see that the spin parts are completely separable\footnote{
Separable means that we can construct the action in terms of summands
for the spin parts and other parts that are associated with the remainder
with certain separate properties we do not specify here.}.
\begin{eqnarray}
\label{Eq::HamTot}
 \HAM{tot}{}&=&
    \frac{p_r^2}{2}-\frac{1}{\rel} +\frac{\Ang^2}{2\rel^2}
 \neanl && 
   +\cInv{2}
   \left\{\frac{\Ang^4 (3 \eta
   -1)}{8
   \rel^4}+\frac{\Ang^2 (3
   \eta -1) p_r^2+2}{4
   \rel^2}-\frac{\Ang^2 (\eta
   +3)}{2 \rel^3}+\frac{1}{8} (3
   \eta -1) p_r^4-\frac{(2 \eta +3)
   p_r^2}{2
   \rel}\right\}
\neanl && 
 +\frac{\cInv{3}}{\rel^3} 
  \Biggl\{
\frac{\left( -J^2+L^2+\SpinTot^2\right)
   \left(12 \epsilon 
   \left(S_2^2-S_1^2\right)+\SpinTot^2 \left(4 \epsilon
   ^2-7\right)\right)}{16 \SpinTot^2} 
   -\underline{
    \underline{ 
   \frac{3 \epsilon  \sin (\phiSpin)
   \sqrt{
   {F_4}(J,L,\SpinTot)
   {F_4}(S_1,S_2,\SpinTot)}}{4 \SpinTot^2}
   }}
   \Biggr\}
 \neanl && 
   +\cInv{4}
   \Biggl\{ 
   \frac{1}{16} (5 (\eta -1) \eta +1) p_r^6 + 
   \frac{\Ang^6}{16 \rel^6} (5 (\eta -1)\eta +1)
   -\frac{\Ang^4}{8 \rel^5} (\eta(3 \eta +20)-5)
   \neanl && 
   +\frac{1}{\rel^4} \left( \frac{3}{16} \Ang^4 (5 (\eta -1) \eta +1) p_\rel^2+\Ang^2 \left(4 \eta +\frac{5}{2}\right) \right)
   +\frac{1}{4 \rel^3} \left( \left(\Ang^2 (5-4 \eta (\eta +5)) p_\rel^2-3 \eta -1\right) \right)
   \neanl && 
   +\frac{1}{\rel^2} \left( \frac{3}{16} \Ang^2 (5 (\eta -1) \eta +1) p_\rel^4+\frac{1}{2} (11 \eta +5) p_\rel^2 \right)
   +\frac{1}{\rel} \left( \frac{1}{8} (5-4 \eta  (2 \eta +5)) p_\rel^4 \right)
   \Biggr\}\,,
\end{eqnarray}
Here, the functions $F_4$ are polynomials of the angular momentum magnitudes
(also see Eqs.~(5.13) and (5.14) of \cite{Tessmer:Steinhoff:Schafer:2013}),
\begin{eqnarray}
F_4(J,L,\SpinTot)&:=&(J-L-\SpinTot) (J+L-\SpinTot) (J-L+\SpinTot) (J+L+\SpinTot) \,, \\
F_4(S_1,S_2,\SpinTot)&:=&(S_1-S_2-\SpinTot) (S_1+S_2-\SpinTot)
   (S_1-S_2+\SpinTot) (S_1+S_2+\SpinTot)\,.
\end{eqnarray}
The doubly underlined sin-term is an oscillatory term {\em for the quasi-circular case only}
in the sense that, as one inserts the solution to the rest
of the Hamiltonian, its average over one time period\footnote{
``Time period'' is as valid as the term ``period'' alone because
it holds $\phiSpin(t-t_0) = \Omega_S \, t$ having $\Omega_S={\rm const.}$} of
$\phiSpin$ is exactly zero.
In the following, we will show how to include the sin-term 
(as a small deviation from the equal-mass limit) into the
equal-mass solution (as the unperturbed problem).
First, we have to find action
and angle variables for the unperturbed problem. Secondly,
with the help of these variables, we perform a canonical
transformation that shifts the sin-term to the order
$\order{\epsilon}{2}$ of the mass difference parameter.

Structurally, this looks as follows:\\

\begin{tabularx}{15cm}{l  l  c}
$\bullet$	& decomposition: 			& $H$ = $\underbrace{\Hint}_{\rm integrable}
                              + \underbrace{\Hpert_{\rm SO}}_{\rm small}$ 
\qquad with \qquad $\order{ \Hpert_{\rm SO} }{}=\epsilon^1$.
\\
\\
$\bullet$	& find AA variables	$(\boldsymbol{I},\boldsymbol{w})$	& $\Hint = \Hint(\boldsymbol{I})$ \qquad $\Rightarrow$ 
\qquad $\Hpert_{\rm SO} = \Hpert_{\rm SO}(\boldsymbol{I},\boldsymbol{w})$ \\
		& for $\Hint$ only:			& \\
\\
$\bullet$	& find generator: & $\Hint(\boldsymbol{I}) \rightarrow \Hint{}^{(1)}(\boldsymbol{I}')$;\qquad
	$\Hpert_{\rm SO}(\boldsymbol{I}, \boldsymbol{w}) 
	\rightarrow	\Hpert_{\rm SO}{}'(\boldsymbol{I}',\boldsymbol{w}') $ \qquad
with	\qquad $\order{\Hpert_{\rm SO}{}'}{}=\epsilon^2$ \,.
\end{tabularx}\\
\vspace{0.5cm}
%

\noindent
To provide more details for finding the AA variables first, we use the full-separation ansatz
for the action $S$ and the function $W$, namely Eq.~(\ref{Eq::DefAction}), 
where $E$ is the energy of the system which is negative in
the bound-orbit case and $|E|$ the value of the binding
energy that appears in the solution for the orbital elements
$a_r, e_r$ and so on; $q$ are
all the spatial coordinates, $q=\{\rel, \phi, \Upsilon, \phiSpin,... \}$.
We justify this separation ansatz below.
The form of $W$ reads
\begin{equation}
 W(q) := W_\rel(\rel) + W_\phi (\phi) + W_\Upsilon (\Upsilon)
    +W_{\rm spin}(q_{\rm spin}) \,,
\end{equation}
where
\begin{equation}
 W_{\rm spin}(q_{\rm spin})
		:= W_{\alpha_{S1}} (\alpha_{S1})
		  +W_{\alpha_{S2}} (\alpha_{S2})
		  +W_{\phiSpin} (\phiSpin) \,.
\end{equation}
Here, the $\alpha_{Sa}$ with $a\in \{1,2\}$ are intrinsic
rotation angles of the individual objects that do not appear
explicitly in the Hamiltonian because of the absence of
spin-spin and spin-squared interaction terms.
The following discussion shows some details of the computation for
the case $\HAM{N+SO}{~no \, \phiSpin}:=\HAM{PM}{N} 
+ \HAM{SO}{LO, \, no\,\phiSpin}$
without the $\phiSpin$-dependent part (the integrable part is what
then remains), as we move to coordinates in which the 3-component
of the orbital angular momentum $\vAng$ is eliminated and only the
scalar contribution $\Ang$ appears\footnote{This procedure can also
be performed in general spherical coordinates where the elimination
has not been done so far. Such a discussion for the Newtonian
case alone can be found in the books~\cite{Goldstein:1981} and
\cite{Vinti:1998}.}.
The extension to the 2PN Hamiltonian without the $\phiSpin$ part
is done in the same way and gives the same structure of terms. One
also observes that the Hamiltonian does not depend on orientations
such as $\Upsilon$, $\phiSpin$, and as mentioned $\alpha_{S1}$ and
$\alpha_{S2}$, which
means that the ``old momenta'' $\Ang, \SpinTot, S_1$ and $S_2$ are
conserved and transformed into themselves (this part of the
generating function being the identity transformation). One can
therefore still
write $\SpinTot$ and $J$ instead of $W_{\phiSpin}(\phiSpin)$ and
$W_\Upsilon (\Upsilon)$, respectively:
\begin{equation}
 \HAM{N+SO}{\rm no \, \phiSpin} = \frac{1}{2}\left( p_r^2 + \frac{\ang^2}{r^2}\right) - \frac{1}{r}
             + \frac{1}{c^3 r^3} \left( J^2 - L^2 - \SpinTot^2 \right)
             \left( 2\eta+3 \right) \,.
\end{equation}
The Hamiltonian does not depend on the variable $\phi$
either and thus one can write down $W_\phi (\phi)=\Ang\,\phi$.
We may write down the above integrable part with the
input of Eq.~(\ref{Eq::DefAction}) and obtain
\begin{equation}
W_\rel'^2 + \frac{1}{\rel^2} \Ang^2 - \frac{2}{\rel}
+ \frac{2}{c^3 r^3} \left( J^2 - L^2 - \SpinTot^2 \right)
             \left( 2\eta+3 \right) = 2 E \,,
\end{equation}
(a prime in $W_\rel'$ means partial derivative 
with respect to $\rel$)
from where one (formally) easily extracts the $W_\rel$ part as
an integral over a square-root. The explicit computation of
the $W_\rel$ part is discussed in Appendix~\ref{SubSec::Integral_of_pr}.

\subsubsection{Results}
The spin-orbit Hamiltonian yields the following action variables,  
\begin{subequations}
\begin{align}
\label{Eq::J_r}
 J_r &=
-\Ang-\frac{1}{\sqrt{2} \sqrt{\NRG}}
+\cInv{2}
   \left(\frac{3}{\Ang}-\frac{(\eta -15) \sqrt{\NRG}}{4
   \sqrt{2}}\right)
\neanl & 
-\frac{\cInv{3} {\JLSGSum} \left(3 \sqrt{1-4 \eta }
   (S_1-S_2) (S_1+S_2)+(2 \eta +3)
   \SpinTot^2\right)}{8 \Ang^3 \SpinTot^2}
\neanl & 
-\frac{\cInv{4} \left(\sqrt{2}
   \Ang^3 (3 \eta  (\eta +10)+35) \NRG^{3/2}+96
   \Ang^2 (5-2 \eta ) \NRG+80 (2 \eta -7)\right)}{64
   \Ang^3}
\,,\\
 J_\phi &=	\Ang \,,\\
 J_\Upsilon &=		J \,,\\
 J_S &=		\SpinTot \,,
\end{align}
\end{subequations}
where the subscripts on the left hand sides
denote the coordinate over which has been integrated,
with the exception of the subscript $S$ for the
spin part for reasons of beauty
\footnote{We assume that the orbital angular momentum
$\vAng$ fulfils $L_z = \vct{e}_z \cdot \vAng >0$, see
\cite{Tessmer:2009} for details.}.
We observe
\begin{equation}
 \left( J_r + J_\phi \right)^2 - a_r = \order{c}{-2}\,.
\end{equation}
Within perturbation theory, the Hamiltonian (energy) can be expressed as
\begin{eqnarray}
\label{Eq::HamiltonianAAVars}
-\Hint(\boldsymbol{J}) &=&
\frac{1}{2(J_r + J_\phi)^2}
\Biggl[
1+ \frac{\cInv{2} }{\left( J_r+J_\phi \right){}^2} \left( \frac{\eta + 9 }{4} + \frac{6 J_r}{J_\phi} \right)
\neanl && 
- \frac{\cInv{3}}{ (J_r+J_\phi) }
\biggl(
\frac{{(J_\Upsilon^2 - J_\phi^2 - J_S^2)} \left(3 \sqrt{1-4
   \eta } (S_1-S_2)
   (S_1+S_2)+(2 \eta +3)
   J_S^2\right)}{4 J_\phi^3
   J_S^2 } 
\biggr)
\neanl && 
+
\frac{\cInv{4}}{\left( J_r+J_\phi \right){}^4} 
\left(
   \frac{5 (7-2 \eta ) J_r^3}{2
   J_\phi^3}+\frac{3 (53-10 \eta ) J_r^2}{2
   J_\phi^2}-\frac{9 (\eta -6)
   J_r}{J_\phi}+\frac{1}{8} ((\eta -7) \eta
   +81)
   \right)
\Biggr]\,.
\end{eqnarray}
We see that $J_\Upsilon$ does not appear in the point-mass parts, and through
Newtonian order only, $J_r$ and $J_\phi$ are degenerate.
We next see what happens when we examine a removal of possible
degeneracies, i.e. a transformation to variables that absorb
conditions of degeneracy.

\subsection{Degenerated systems: Delaunay variables for the  spin-orbit Hamiltonian}
If a system of n degrees of freedom has an m-fold degeneracy, meaning that
the first $m$ frequencies are not linearly independent in the sense
\begin{equation}
 \sum_{i=1}^{m} n_{\alpha i} \omega_i=0 \,,
 \qquad  \text{$\alpha$:~labeling~the}~\alpha^{\text{th}}~\text{condition}\,,
\end{equation}
one can construct a generator of type 2 -- in the sense of common
literature on theoretical mechanics --\newline
$F_2$: ($\omega\,\rightarrow\,\bar{\omega},~J \rightarrow I$) of the form
\begin{equation}
 F_2\left( \omega_i, I_i \right)
 = \sum_{k=1}^{m} \sum_{i=1}^{n} n_{ki} \omega_i I_k
  +\sum_{k=m+1}^{n} \omega_k I_k
\end{equation}
where $n_{ki}$ is a coefficient of the $k^{\rm th}$ degeneracy
condition to connect the angle variables with index $i$,
such as for a fictitious set of variables $\boldsymbol{\omega}^*$
\begin{subequations}
\begin{align}
k=1: \qquad n_{11} \boldsymbol{\omega}^*_1 + n_{13} \boldsymbol{\omega}^*_3 + \dots = & \, 0 \,, \\
k=2: \qquad n_{21} \boldsymbol{\omega}^*_1 + n_{22} \boldsymbol{\omega}^*_2 + \dots = & \, 0 \,, \\
\text{\dots and} \text{ so }& \text{on,}\tag{3.23n}
\end{align}
\end{subequations}
resulting in
\begin{align}
\label{Eq::F2_degeneracy}
\begin{tabular}{lll|r l}
 $\bar{\omega}_k$ 	&=& $\frac{\partial F_2}{\partial I_k}$
 =$\begin{cases}
     \sum_{i=1}^{n}n_{ki}\omega_i	& \text{for~} k=  1,...,m\\
    \hfill \omega_k			& \text{for~} k=m+1,...,n
   \end{cases}$ 
   & ~~~$\bar{\omega}_k = \frac{\partial H}{\partial I_k} = 0$ for k=1...m \\
&&&& \\
 $J_i$			&=& $\frac{\partial F_2}{\partial \omega_i}$
 =$\sum_{k=1}^{m} n_{ki} I_k + \sum_{k=m+1}^{n} \delta_{ki} I_k $
 & $\Rightarrow H=H(\boldsymbol{I})$
\end{tabular}
\end{align}
In the case of a Newtonian binary compact object we observe%
\footnote{Newtonian binaries do not suffer periastron shift,
therefore the radial period is the same as the angular, see
Eq.~(\ref{Eq::HamiltonianAAVars}).}
\begin{equation}
 \omega_r - \omega_\phi = 0 \quad \Rightarrow \quad
 n_{1\phi} = -1, \quad n_{1r}=1, \quad n_{1\Upsilon}=0;
\end{equation}
so our  generating function will look as follows,
\begin{equation}
 F_2 \left( \boldsymbol{\omega}, \boldsymbol{I} \right)
 = \left( \omega_\phi - \omega_r \right) I_1
   +\omega_r I_2
   +\omega_\Upsilon I_3
   +\omega_S I_4 \,.
\end{equation}
From Eq.~(\ref{Eq::F2_degeneracy}) the transformation of the momenta and coordinates yields
\begin{align}
\begin{tabular}{ l l l}
 $ J_r		$ &= $ I_2 - I_1 \,, $	\qquad \qquad 	& $w_1 = w_\phi - w_r \,, $\\
 $ J_\phi	$ &= $ I_1 \,,$ 			& $w_2 = w_r \,, $\\
 $ J_\Upsilon	$ &= $ I_3 \,,$ 			& $w_3 = w_\Upsilon \,, $\\
 $ J_S		$ &= $ I_4 \,,$ 			& $w_4 = w_S \,, $\\
\end{tabular}
\end{align}
\noindent
The transformation from old to new momenta is simply obtained
by inversion of the above system.
The total integrable Hamiltonian $\Hint$, written in terms of the new $\boldsymbol{I}$, then reads
\begin{eqnarray}
- \Hint\left( \boldsymbol{I} \right) &=&
\frac{1}{2 I_2^2}
+\frac{\cInv{2}}{2 I_2^2} \left( \frac{\eta -15}{4 I_2^2} + \frac{6}{I_1 I_2} \right)
+\cInv{3}\frac{\left(I_1^2-I_3^2+I_4^2\right)
   \left((2 \eta +3) I_4^2+3 \sqrt{1-4\eta } (S_1-S_2)
   (S_1+S_2)\right)}{8 I_1^3
   I_2^3 I_4^2}
\neanl && 
+\frac{\cInv{4}}{2 I_2^2}
\left(
\frac{5 (7-2 \eta )}          {2 I_1^3 I_2^{}}
   +\frac{3(4 \eta -35)}      {2 I_1^{} I_2^3}
   +\frac{(\eta-15) \eta +145}{8 I_2^4}
   +\frac{27}                 {  I_1^2 I_2^2}
\right)\,.
\end{eqnarray}
This is the integrable part as a function of what is known as Delaunay variables
$\left(I_1, I_2, I_3; \omega, \ell, \Upsilon \right)$ and their extension of
the spin magnitudes and total angular momentum, see below for explanation.
Taking the action variables in Ref.~\cite{Damour:Schafer:1988} 
which differ by the re-definition 
$I_2\rightarrow i_3$,
$I_1\rightarrow i_2$,
this exactly reproduces those
authors' result through 1PN, see their Eq.~(3.13).
Again, this labelling discrepancy results from the missing degeneracy
in our Hamiltonian that would be present if we used an unspecified\footnote{
``Unspecified'' means that we would take a general direction of $\vct{L}$
and look for the spherical coordinate contributions,
not only the planar problem in the unperturbed 
Newtonian case.}
frame for a derivation instead.
To make contact with Vinti's notation \cite{Vinti:1998} of Delaunay's variables,
marked with subscript ``$D$'', let us give the
following (Newtonian) relations, which will be needed
for the first-order perturbation generator:\\
\begin{eqnarray}
\label{Eq::NewtonianDelaunays}
 L_D &= \sqrt{a} 		=I_2\,, \qquad \qquad	& \ell_D = \MeMo \left(t+\beta_1 \right)\,, \\
 G_D &= \left| \Ang \right|	=I_1\,, \qquad \qquad & {g_D=\omega}\,, \\
 H_D &= L_Z+S_Z =J_\Upsilon=I_3 \qquad \qquad & h_D=\Upsilon\,, \\
 S_D &= \SpinTot \qquad \qquad & s_D = \phiSpin\,, \\ \hline
 \Sigma_{1D} &= S_1 \qquad \qquad & \sigma_{1D} = \alpha_{s1}\,, \\
 \Sigma_{2D} &= S_2 \qquad \qquad & \sigma_{2D} = \alpha_{s2}\,,
\end{eqnarray}\\
\noindent
$\beta_1$ being the linear-in-time coordinate function that is 
associated with the constant ``momentum'' $\alpha_1$, and $\omega$
as the argument of the pericenter.
The $\left(\sigma_{aD}, \Sigma_{aD} \right)$ section of the above block of
variables is not present in the Newtonian case and has been added to
complete the phase space. 
That means that the variable $\phi$ has been removed by
means of the Newtonian degeneracy condition.
Taking Eqs.~(3.14) and (3.15) of \cite{Damour:Schafer:1988},
\begin{eqnarray}
 \MeMo		&:=& \frac{\partial \Hint}{\partial I_2} = \frac{\partial \Hint}{\partial L_D} \,, \\
 k\,\MeMo	& =& \frac{\partial \Hint}{\partial I_1} = \frac{\partial \Hint}{\partial G_D} \,,
\end{eqnarray}
we obtain for the periastron advance parameter $k$
\begin{equation}
 k := \frac{\Phi-2\pi}{2\pi} = \frac{3}{c^2 I_1^2} = \frac{3}{c^2 \Ang^2} + \order{c}{-3}\,,
\end{equation}
which is a well-known result.
Further, the relations $e	= \sqrt{1-2E\Ang^2}$
and $E = \frac{-1}{2a}$ hold
-- again only in the Newtonian case -- 
such that
\begin{equation}
\label{Eq::DelaunayEccentricity}
 e^2 = 1 - \left(\frac{G_D}{L_D}\right)^2 +\order{c}{-2}\,.
\end{equation}
We are aware that in the quasi-circular limit, $G_D$ and $L_D$ are degenerate.
In that case, one is forced to transform to another set of variables that incorporates
this degeneracy, for example 
the Poincar\'e elements as pointed out in the notes of Howison and Meyer \cite{Howison:Meyer:2011}
or to the approach in \cite{Tessmer:Steinhoff:Schafer:2013}.
However, our calculation starts from the eccentric case, meaning that the startup
to the solution is not evaluated on the circular orbit. One can deal generally
with an eccentric system and let, finally on the solution level, let $e$ tend to zero.

Subsequently, we will present basics of canonical perturbation theory
for the action and angle variables and the application to the eccentric
spin-orbit problem.
This perturbation theory aims to find a generator for a canonical transformation
that shifts contributions of the total interaction Hamiltonian which have
oscillatory dependencies on phase space coordinates (called $\boldsymbol{\phi}$ here)
to a higher order of the
small expansion parameter $\epsilon$, resulting in a new Hamiltonian that
only depends on the transformed momenta.
As for the circular-orbit case in
Ref.~\cite{Tessmer:Steinhoff:Schafer:2013}, the mass difference function
will be chosen to be the mentioned smallness parameter.

\section{Hamilton-Jacobi perturbation theory with action-angle variables}
\label{Sec::HJ_Perturbation_Theory}
As it could be seen in Eq.~(\ref{Eq::HamTot}), the 
Hamiltonian contains a term that depends on the spin orientation phase
$\phiSpin$ which is of the order $\order{\epsilon}{1}$ and not included
in the known solutions.
Writing the total Hamiltonian in terms of the Delaunay elements in which
the unperturbed Hamiltonian only depends on the momenta and with the help
of a further canonical transformation, we
like to shift that expression to order $\order{\epsilon}{2}$.
Below, we list the basic properties of such a general canonical
transformation. \newline

The task is to solve the Hamilton-Jacobi equation
\begin{equation}
 H \left( \boldsymbol{\phi}, \partial_{\boldsymbol{\phi}} S\left(\boldsymbol{\phi}, \boldsymbol{I}' \right)\right)
 =
 H'\left( \boldsymbol{I}' \right)\,,
\end{equation}
where the right hand side only depends on the new momenta $\boldsymbol{I}'$, not on the angles $\boldsymbol{\phi}$,
perturbatively, although the existence of a solution $S$ may not be guaranteed.
We expand the generator $S$ around the identity transformation
($\boldsymbol{\phi} \rightarrow \boldsymbol{\phi}, \boldsymbol{I} \rightarrow \boldsymbol{I}'=\boldsymbol{I}$)
in powers of the perturbation parameter $\epsilon$
and set
\begin{eqnarray}
 S  \left( \boldsymbol{\phi}, \boldsymbol{I}' \right) &=& S_0 
 + \epsilon   S_1 \left( \boldsymbol{\phi}, \boldsymbol{I}' \right)
 + \epsilon^2 S_2 \left( \boldsymbol{\phi}, \boldsymbol{I}' \right)
 + \order{\epsilon}{3} \,, \\
 S_0\left( \boldsymbol{\phi}, \boldsymbol{I}' \right) &=& \boldsymbol{\phi} \cdot \boldsymbol{I}' \qquad \text{(=identity transformation)}\,, \\
 S_1\left( \boldsymbol{\phi}, \boldsymbol{I}' \right) &\dots& \text{ to be found,}
\end{eqnarray}
which gives, up to first order in $\epsilon$, 
\begin{eqnarray}
 H_0 \left( \partial_{\boldsymbol{\phi}}(S_0 + \epsilon S_1) \right)
   + \epsilon H_1 \left(\boldsymbol{\phi}, \partial_{\boldsymbol{\phi}}(S_0 + \epsilon S_1) \right)
 &=& H' (\boldsymbol{I}') \,, \nonumber \\
 H_0 \left( \boldsymbol{I}' \right) 
 + \epsilon \partial_{\boldsymbol{I}} H_0 \left( \boldsymbol{I} \right)|_{\boldsymbol{I}=\boldsymbol{I}'} \partial_{\boldsymbol{\phi}}(S_1)
 + \epsilon H_1 \left(\boldsymbol{\phi}, \boldsymbol{I}' \right)
 &=& H' (\boldsymbol{I}') \,,
 \end{eqnarray}
where we have used $H_0=H_0(\boldsymbol{I})$.
Subtracting $H' (\boldsymbol{I}')$ gives zero on the right hand side.
The resulting relation can be fulfilled only if each coefficient
of powers of $\epsilon$ is equivalent to 0, i.e.
\begin{eqnarray}
0:& H_0 (\boldsymbol{I}') 
&\stackrel{!}{=} H'(\boldsymbol{I}') \,,\\
\label{Eq::Generator_cond_1stOrder}
1:& \partial_{\boldsymbol{I}} H_0 \left( \boldsymbol{I} \right)|_{\boldsymbol{I}=\boldsymbol{I}'} \partial_{\boldsymbol{\phi}}(S_1)
&\stackrel{!}{=} -H_1 \left( \boldsymbol{\phi}, \boldsymbol{I}' \right) \nonumber \\
\Rightarrow &
\boldsymbol{\omega} \cdot \partial_{\boldsymbol{\phi}} S_1 \left( \boldsymbol{\phi}, \boldsymbol{I}' \right) & = -H_1 \left( \boldsymbol{\phi}, \boldsymbol{I}' \right)
\end{eqnarray}
Here, $\boldsymbol{\omega}=\boldsymbol{\omega}(\boldsymbol{I}')$ holds.%
\footnote{The $\boldsymbol{\omega}$ in above equation
is computed using the derivative with respect to the
unprimed variables, thereafter replacing all
variables $\boldsymbol{I}$ by the primed ones $\boldsymbol{I}'$
\emph{without using the variable transformation}
which, anyway, still has to be obtained.}
To solve this, we make the Fourier ansatz
\begin{equation}
\label{Eq::Pert_Fou_Ansatz}
 S_1 \left( \boldsymbol{\phi} \right)
 = \sum_{k_1, \dots k_j = -\infty}^{\infty} S_{1 \boldsymbol{k}} \, 
   e^{ i\, \boldsymbol{k} \cdot \boldsymbol{\phi} }
\end{equation}
and accordingly for $H_1$.
Here, $\boldsymbol{k}$ is a multi-index, and 
$\boldsymbol{k} \cdot \boldsymbol{\phi} = \sum_{i} k_i \phi_i$.
With this input and Eq.~(\ref{Eq::Generator_cond_1stOrder}) we obtain
\begin{eqnarray}
 \partial_{\phi_j} S_1(\boldsymbol{\phi})	
 &=& i\, \sum_{ \boldsymbol{k} } S_{1\boldsymbol{k}}\, k_j e^{ i\, \boldsymbol{k} \cdot \boldsymbol{\phi} } \,, \\
 \sum_{ \boldsymbol{k} } \Bigl\{
 i \, \boldsymbol{\omega} \cdot \boldsymbol{k} \, S_{1\boldsymbol{k} } (\boldsymbol{I}') 
 + H_{1\boldsymbol{k}}(\boldsymbol{I}')
  \Bigr\}e^{ i\, \boldsymbol{k} \cdot \boldsymbol{\phi} }	&=& 0 \,.
\end{eqnarray}
This has to hold for arbitrary $\boldsymbol{\phi}$, meaning that
all the $\boldsymbol{k}$-coefficients vanish:
\begin{equation}
\label{Eq::generator_with_maybe_zero_denominator}
 S_{1 \boldsymbol{k}} \left( \boldsymbol{I}' \right) = i \frac{ H_{1\boldsymbol{k}} \left( \boldsymbol{I}' \right) }
 {\boldsymbol{\omega} \cdot \boldsymbol{k} } \,.
\end{equation}
The reader should be warned that the inner product
in Eq.~(\ref{Eq::generator_with_maybe_zero_denominator})
may vanish for special systems. We have to show that,
for our problem, the system does not fulfil any exact
degeneracy condition.
The Hamiltonian $H_0$ will be identified
with $\Hint$ and $H_1$ with $\Hpert$ in the subsequent
sections.

\subsection{The perturbing Hamiltonian: Series expansion around the circular and equal-mass case}

In this section we extract the oscillatory parts
of the full spin-orbit problem.
In \cite{Tessmer:Steinhoff:Schafer:2013} we saw that there exist
oscillatory terms for the circular orbit case. In addition, eccentricity
will also create oscillations. Therefore, we expand the full Hamiltonian
around the equal-mass case (here: to the first power of $\epsilon$) and, as well, 
present it in powers of eccentricity $e$ through fourth order.

The sin-part of the Hamiltonian symbolically reads
 \begin{eqnarray}
 \HAM{SO}{pert} = \epsilon \frac{1}{r^3} {\cal G} (\mathbf{X}_{\rm ang}) \sin {\phi_S} \,,
 \end{eqnarray}
where ${\cal G}$ is a function of the angular momenta amplitudes $\mathbf{X}_{\rm ang}$ solely,
see Eq.~(\ref{Eq::HamTot}). Our task is now
to express this Hamiltonian in terms of the Delaunay variables from the previous section
with the help of an eccentricity expansion around the initial solution.
The ``solution'' to the unperturbed problem will be that for the case
$m_1 = m_2$ and $e \ne 0$, see Eqs.~(\ref{Eq::QKP_EqualMass})~\cite{Konigsdorffer:Gopakumar:2005}.

\noindent
We may lend help from Ref. \cite{Tessmer:Schafer:2010}, where
inverse powers ($n$) of $r$ are expanded first in harmonics of
$\EccAno$ and afterwards 
in $\ell_D$ which is the desired result. We already know that
it holds for $A:=\left(1-e \cos \EccAno \right)$
\begin{eqnarray}
 A^{-n} = \sum_{j \ge 0} {\cal A}_j^{(n)} \cos (j \ell_D)
\end{eqnarray}
where ${\cal A}_j^n$ is a relatively complicated function of the eccentricity,
factorials and Bessel functions of the summation index $j$
(also see Eq.~(37) of \cite{Alexander:1987}, also the standard 
material in \cite{Watson:1980} and, for further investigations
on a post-circular expansion for gravitational wave generation
in the Newtonian case, Ref.~\cite{Yunes:Arun:Berti:Will:2009}).
We may expand ${\cal A}^{-n}$ to, say, fourth order%
\footnote{A general expression for arbitrary $n$ seems to be
obtainable, but has not been found yet. Its coefficients for finite $n$ are easy
to be calculated manually.}
in $e$:
\begin{eqnarray}
\label{Eq::A_of_inverse_n}
 A^{-n} &=&
   1
   +e^2
   \left(\frac{n^2}{4}-\frac{n}
   {4}\right)
   +e^4
   \left(\frac{n^4}{64}+\frac{n
   ^3}{32}-\frac{n^2}{64}-\frac{n}{32}\right)
+\cos (\ell_D) \left(e^3
   \left(\frac{n^3}{8}+\frac{n^
   2}{8}-\frac{3 n}{8}\right)
   +en\right)
\neanl &+&   
   \cos (2 \ell_D) \left(e^4
   \left(\frac{n^4}{48}+\frac{n
   ^3}{8}-\frac{n^2}{48}-\frac{11
   n}{24}\right)+e^2
   \left(\frac{n^2}{4}+\frac{3
   n}{4}\right)\right)
   +e^3 \cos (3 \ell_D)
   \left(\frac{n^3}{24}+\frac{3
   n^2}{8}+\frac{17
   n}{24}\right)
\neanl &+&
   e^4 \cos (4 \ell_D) \left(\frac{n^4}{192}+\frac{3
   n^3}{32}+\frac{95
   n^2}{192}+\frac{71
   n}{96}\right)
\end{eqnarray}
to read-off the coefficients ${\cal A}_j^{(n)}$.
{\sl {\bf An important remark:} This has been done to Newtonian order only.
A generalisation including 1PN terms in the perturbing function would
let us distinguish the "radial" and "time"
eccentricities $e_r$ and $e_t$
appearing in the solution $\ell_D(\EccAno)$ and the
expression $A(r)$ to be combined in an extension of our
Eq.~(\ref{Eq::A_of_inverse_n}).
Going further to 2PN order would mean to include
Eqs.~(102)--(110) of Ref. \cite{Tessmer:Schafer:2011}
and an expansion of regularised hyper-geometric functions
to some order of $e_t$. Our aim is to deliver the knowledge for the leading
order, so we sketch the way for the Newtonian Kepler equation only.}

%

\subsection{Examining the perturbation and the generator in the Fourier domain}
With these inputs, we can easily express the perturbing Hamiltonian $\HAM{SO}{pert}$
as
\begin{equation}
\HAM{SO}{pert}
=
\frac{1}{L_D^6}
\epsilon {\cal G} (\mathbf{X}_{\rm ang}) \,
\sin (\phiSpin) \,
\sum_{j=0}^{\infty}
\mathcal{A}^{(3)}_j\left( G_D, L_D \right) \cos (j\ell_D)\,.
\end{equation}
In expanded and full-canonical form (except of the Newtonian $e$ appearance which
can be avoided by using Eq.~(\ref{Eq::DelaunayEccentricity})), it reads
\begin{eqnarray}
 \HAM{SO}{pert} &=& -\frac{3 \cInv{3} \epsilon  \sin
   (\phiSpin)}{4 \Ang_D^6
   \SpinTot^2}
   \Biggl\{
   1
   +\frac{3 e^2}{2}
   +\frac{15 e^4}{8}
   +\left(\frac{27 e^3}{8}+3e\right) \cos (\ell_D)
   +\left(\frac{7 e^4}{2}+\frac{9e^2}{2}\right) \cos (2 \ell_D)
   \neanl && 
   +\frac{53}{8} e^3 \cos (3 \ell_D)
   +\frac{77}{8} e^4 \cos (4 \ell_D)
   \Biggr\}
   \sqrt{F_4(J,{\Ang},\SpinTot)
   F_4(S_1,S_2,\SpinTot)}\,.
\end{eqnarray}
Therefore, we need Fourier transformations of
\begin{subequations}
 \begin{align}
  {\cal F}_k(\sin(a \phiSpin), \phiSpin) &= \frac{1}{2\mathrm{i}} 
  \left(\delta_{a,k}-\delta_{a,-k} \right) \,, \\
  {\cal F}_k(\cos(a \ell_D), \ell_D)   &= \frac{1}{2} 
  \left(\delta_{a,k}+\delta_{a,-k} \right) \,,
 \end{align}
\end{subequations}
with integer $k$, where ${\cal F}(f(q),q,k)$ is the $k^\text{th}$ Fourier coefficient
of the function $f$,
\begin{eqnarray}
  		f(q) &=&
     \sum_{j=-\infty}^{\infty} 
     {\cal F}_j(f(q), q) e^{\mathrm{i} j q} \,,
\\
{\cal F}_k(f(q),q)	&:=&	\frac{1}{2\pi}\displaystyle \int_{0}^{2\pi} f(q) \, 
\text{exp} \left(-\mathrm{i} k q \right) {\rm d}q \,.
\end{eqnarray}
As the zero$\text{th}$-order Hamiltonian is independent of $\omega_j$,\
defining the new main angular velocities $\omega_i$ with respect to the
Delaunay variables according to
\begin{equation}
 \omega_j = \left. \frac{\partial H_0 (\boldsymbol{I})}{\partial I_j}
            \right|_{\boldsymbol{I}=\boldsymbol{I}'} \,, \qquad j=1, \dots, 4 \,,
\end{equation}
yields
\begin{eqnarray}
\label{Eq::Generator_FouDom}
 S_{1\mathbf{k}} \left( \boldsymbol{I}' \right) &=& 
\frac{
1
}{ \scpm{\mathbf{k}}{\boldsymbol{\omega}} }
\frac{1}{2}
\frac{\cInv{3} \epsilon}
     {L_D^6   S_D^2}
\sqrt{ F_4(J,\Ang,\SpinTot) F_4(S_1, S_2, \SpinTot) }
\left(\delta _{1,-k_4}-\delta_{1,k_4}\right) \times
\neanl &&
\Bigl(
 \frac{3}{4}
 +\frac{9}{8}  e   \left(    \delta _{1,-k_1}+   \delta_{1,k_1}\right)
\neanl &&
 +\frac{9}{16} e^2 \left(  3 \delta_{2,-k_1} + 3 \delta_{2,k_1}+2\right)
\neanl &&
 +\frac{3}{64} e^3 \left( 27 \delta _{1,-k_1}+27 \delta_{1,k_1}
               +53 \left(    \delta_{3,-k_1} +   \delta_{3,k_1}\right)
                  \right)
\neanl &&
  \left.
 +\frac{3}{64} e^4 \left( 28 \delta_{2,-k_1} +28 \delta_{2,k_1}
                         +77 \delta_{4,-k_1} +77 \delta _{4,k_1}+30\right)
\Bigr)
\right|_{ \boldsymbol{I} \rightarrow \boldsymbol{I}'}
+\order{e}{5}
\end{eqnarray}
in the Fourier representation analogous to
Eq.~(\ref{Eq::Pert_Fou_Ansatz}).\footnote{The reader
is reminded of the fact that this relation is not
a time-Fourier representation.} 

The solution to the perturbed problem now consists
of performing the coordinate transformation explicitly;
that means a transformation to the new momenta and new
phase coordinates.
Since the generator depends on the old $\boldsymbol{\phi}$
and new $\boldsymbol{I}'$, use has to be made of the relations
\begin{eqnarray}
\boldsymbol{I} &=& 
\frac{\partial {S}}
     {\partial \boldsymbol{\phi}} \,,\\
\boldsymbol{\phi}' &=& 
\frac{\partial {S}}
     {\partial \boldsymbol{I}'} \,.
\end{eqnarray}
The first set of equations is to be inverted for
$\boldsymbol{I}'$, then the resulting relations
have to be inserted in the second set to eliminate
$\boldsymbol{I}'$ in favour of the old $\boldsymbol{I}$
to finally obtain
\begin{eqnarray}
\boldsymbol{\phi}' &=& \boldsymbol{\phi}'
(\boldsymbol{\phi},\boldsymbol{I}) \,,\\
\boldsymbol{I}'    &=& \boldsymbol{I}'
(\boldsymbol{\phi},\boldsymbol{I})\,.
\end{eqnarray}
For a second transformation, the full information to the
solution (for the new Hamiltonian in the new coordinates)
has to be found.
The reader should be aware that, going to the n$^{\rm th}$
transformation, {\em all} terms to order n have
to be kept in the generator approximation process
until the end.
That means that also the generating function $S$ itself
has to be Taylor expanded to order $\order{\epsilon}{n}$.

The convergence of the Fourier
series Eq.~(\ref{Eq::Pert_Fou_Ansatz}), also having a hidden dependence on the 
higher-order-in-$e$ contribution that provides the
higher $\mathbf{k}$ terms 
has to be discussed. This can be done with the help
of the \em{Kolmogoroff, Arnold and Moser} (KAM) theory.

\subsection{Some remarks about the non-degeneracy of the Delaunay frequencies \texorpdfstring{$\omega_j$}{\omega}}
The KAM theory {\cite{Wayne:2008}} states that for sufficiently non-degenerate
systems (in classical lectures, other conditions than $\sum n_{\alpha i} \omega_i \neq 0$
are given; the strong nonresonance: the existence of
constants $\alpha>0$ and $\tau>0$ such that 
$|\langle \boldsymbol{k} \boldsymbol{\omega} \rangle| 
\ge
\frac{\alpha}{|\boldsymbol{k}|^\tau}$
for all $0\neq k\in \mathbb{Z}^n$
with $|\boldsymbol{k}| = \sum_{i}|k_i|$),
the series expansion
(\ref{Eq::generator_with_maybe_zero_denominator}) converges.
In Arnold's book \cite{Arnold:1974}, p. 408, the following condition
\begin{equation}
\label{Eq::Arnolds_non_degeneracy}
{\rm det} \left | \frac{\partial^2 H_0}{\partial \boldsymbol{I}^2}  \right | \neq 0 \,,
\end{equation}
was provided which guarantees conservation of most invariant tori
under small perturbations.
We like to state that, although we insert the Newtonian-order
solution into the perturbation generator because of our PN truncation,
what we like to perturb is {not} the Newtonian solution but the
equal-mass 1PN SO + 2PN PM solution. In that context, our approximation
is too crude to see the periastron advance and spin precession effects
in the generator itself, so what will be required in an extension to
higher PN orders of this generator in order to include the ``missing'' dynamics.
Although the non-degeneracy condition (\ref{Eq::Arnolds_non_degeneracy})
is not fulfilled in the Newtonian case (the denominator in 
(\ref{Eq::Generator_FouDom}) then anyway would only contain one single frequency
rather than a summation),
it definitely is so in the PN case. Therefore,
the general relativistic solutions are ``much more stable'' with respect
to perturbations.

It is, for the time being, unclear (i), how large the mass difference and
(ii), how large the binding energy of the system is allowed to be (possibly
generating degenerate frequencies at some point of the evolution downwards inspiral)
before the deformed tori are finally destroyed.

\section{Conclusions}
In this article, we presented a first-order solution to the eccentric two-body
problem with spin-orbit coupling having slightly different masses. We expressed
the solution to the well-known equal-mass solution in terms of Delaunay-type
variables.
With the help of these variables, we constructed 
a canonical transformation which shifts the perturbing Hamiltonian part, characterised 
by the sin-function of the spin orientation phase $\phiSpin$ and being of first order of
the mass-difference function $\epsilon$, to second order where it may also
contain $\cos \phiSpin$ terms.

As a task to remain for a future publication it has to be found out how large
the mass difference is allowed to be before the deformed KAM tori are destroyed.
Further, one has to take into account the next-to leading order of the spin-orbit
interaction, which means that in the Fourier expansion of the inverse distance it
has to be distinguished between $e_t$ and $e_r$, which modifies the solution
at higher orders of inverse $c$.

\noindent
{\bf A remark on Delaunay elements in higher orders of 
$c$:}
We computed the quantity $J_r$ as a definite integral
over the radial variable. In order to express the time
$t$ as a function of the variable $\ell_D$ in higher
orders of $c$, we may use a generating function
of the form
\begin{equation}
W = \ang \phi 
+ J \Upsilon 
+ \int_{r_+}^{r} f_r(r') {\rm d}r'
+ W_{\rm spin}\,,
\end{equation}
where $r_+$ denotes the radial distance at the periastron
and $f(r)$ is constructed in such a way that
the new variable $\ell_D$ is directly related
to the time $t$ as a derivative of $W$
and closely related to the Kepler equation
(see standard texts on Delaunay elements, e.g.
\cite{Vinti:1998}, and 
also the quasi-Keplerian parameterisation for
higher PN orders, for example 
\cite{Wex:1995,
      Memmesheimer:Gopakumar:Schafer:2004,
      Tessmer:Hartung:Schafer:2010}).
Note that $L$ is the orbital angular momentum and to
be distinguished from the energy-related Delaunay 
element $L_D$.
We could use the Newtonian relations from common 
literature (which did not have to be re-calculated)
in the current article, but the above relation
has to be taken care of in a further development.
It may turn out that, therefore, not much effort or new
quality of calculation has to be considered to obtain
the higher PN-order result.

Finally, two more problems are remaining in this arena. The first one is to tackle
spin-squared and spin-spin interactions. Those Hamiltonians have a simple appearance
in the coordinate-independent form, but being expressed in terms of the Delaunay-type (or Hill-type)
variables or those in \cite{Tessmer:Steinhoff:Schafer:2013}, they get complicated
in comparison to spin-orbit interactions.
This circumstance deserves a careful consideration.
The second one is the treatment of radiation reaction, where it is currently
unclear how to combine the radiation interaction terms and the eccentricity
vs. unequal-mass precession in reasonable order for an analytic consideration.

\acknowledgments
The authors wish to thank Professor Manfred Schneider for useful discussions.
Thanks also go to two anonymous referees for giving useful hints for improving
the paper.
This work was funded by the Deutsche Forschungsgemeinschaft
(DFG) through SFB/TR7 ``Gravitational Wave Astronomy'' and
the DLR through ``LISA -- Germany''.

\appendix
\section{A short excursion to the contour integration for \texorpdfstring{${p_r}$}{pr}}
\label{SubSec::Integral_of_pr}
\begin{figure}[hc!]
\includegraphics[width=\textwidth, angle=-0,scale=0.70]{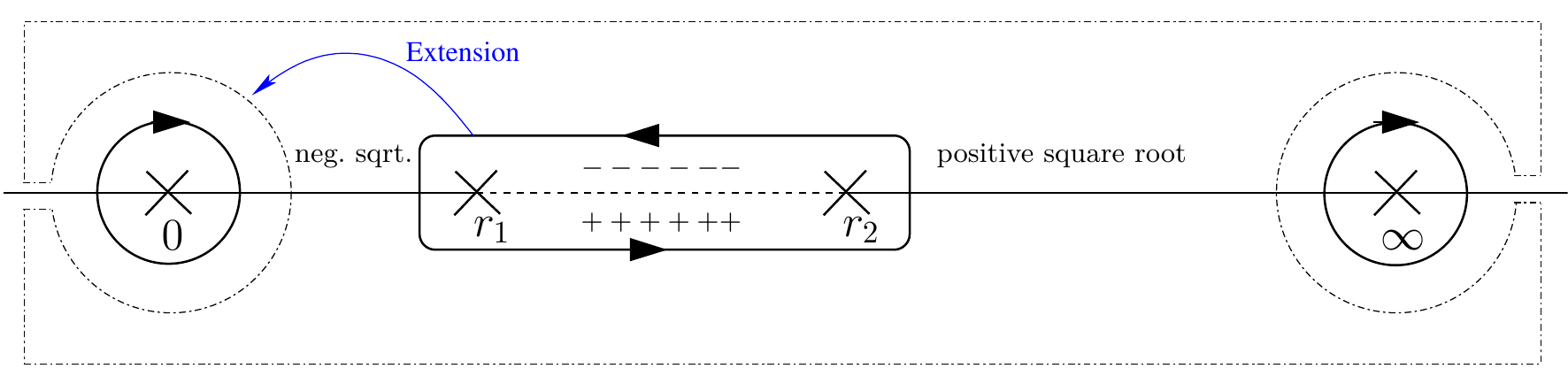}
\caption{
Contour integral (also see Refs. \cite{Goldstein:1981} 
and \cite{Sommerfeld:1951})
for the application of the method of residues.
The values $r=0$ and $r$\,``=''$\infty$ are the only singular points.
Below the real axis, the path is towards the apastron $r_2$ and thus 
the square-root has positive sign -- a closed path computation is then
possible. The dashed line is an intermediate step  in deforming the
contour in such a way that $r=0$ and $r=\infty$ are the only excluded
points.
\label{Fig::Contour}}
\end{figure}

The integral for $J_r$ can be computed by applying the method of residues.
The integration is running 
from $r_1$ to $r_2$ and back, defining $r_1$ to be the inner and $r_2$
to be the outer boundary, $r_1<r_2$, see Fig \ref{Fig::Contour}. 
These two points represent the boundary of a branch cut
in the complex r-plane.
On the journey from $r_1$ to $r_2$, $p_r$
is positive, and backwards negative. Thus, a single integration
from $r_1$ to $r_2$ can be split into $\frac{1}{2}$ times an
integration above plus one below the real axis, taking into account
the change of the signs when changing the direction of the path.
What follows is an expansion of the integration to the whole real axis.
There are only 2 singular points, namely $0$ and $\infty$.
The sign of the square-root is ``-'' for $r < r_1$ and is
``+'' for $r>r_2$. Let $f(r)$ denote the radicand in $p_r$. Then
the final result for $J_r$ is (see the rotation directions
and the signs of the radicand to be taken!)
\begin{equation}
 J_r = \frac{1}{2\pi} \times
 2 \pi i \, \left(  {\rm Res} \left( \sqrt{f}, r=0 \right) 
             - {\rm Res} \left( \sqrt{f}, r \rightarrow \infty \right) \right)\,.
\end{equation}


\end{document}